\documentclass[twocolumn,showpacs,preprintnumbers,amsmath,amssymb]{revtex4}

\usepackage[dvips]{graphicx}
\usepackage{subfigure}
\usepackage{dcolumn}
\usepackage{bm}
\usepackage{amsmath}

\begin{document}

\title{Self-organized Emergence of Navigability on Small-World Networks}
\author{Zhao Zhuo$^{1}$}
\author{Shi-Min Cai$^{1}$}
\email{csm1981@mail.ustc.edu.cn}
\author{Zhong-Qian Fu$^{1}$}
\email{zqfu@ustc.edu.cn}
\author{Wen-Xu Wang$^{2}$}
\affiliation{$^{1}$Department of Electronic
Science and Technology, University of Science and Technology of
China, Hefei Anhui, 230026, PR China\\
$^{2}$Department of Systems Science, School of Management,
Center for Complexity Research, Beijing Normal University,
Beijing 100875, China}
\date{\today}

\begin{abstract}
This paper mainly investigates why small-world networks are navigable
and how to navigate small-world networks. We find that the navigability
can naturally emerge from self-organization in the absence of prior
knowledge about underlying reference frames of networks. Through a
process of information exchange and accumulation on networks,
a hidden metric space for navigation on networks is constructed.
Navigation based on distances between vertices in the hidden metric
space can efficiently deliver messages on small-world networks,
in which long range connections play an important role.
Numerical simulations further suggest that high cluster
coefficient and low diameter are both necessary for
navigability. These interesting results provide profound insights
into scalable routing on the Internet due to its distributed and
localized requirements.
\end{abstract}


\maketitle

\section{Introduction}

Small-world (SW) networks are ubiquitous in nature and society.
Travers and Milgram discovered small-world phenomenon through
delivering letters among people in late 1960s \cite{Travers1969}. In
the experiment, each participant could only deliver letters to a
single acquaintance who was more possible to deliver letters to target
persons based on their own judgement. Relying on this greedy routing
strategy, or so-called navigation, at last 29 percent of letters
reached target persons and the average length of acquaintance chains
of letters that were successfully sent was 6. Recent experiments have
proved that the greedy routing strategy could efficiently pass
messages on email networks and online social service networks
\cite{Dotts2003,Adamic2005,Liben-Nowell2005}. These striking results
suggest that people are connected with much shorter chains than our
imagination and they can find the short paths based solely on local
information, regardless of the network size and the topological
distances among people.

Navigability of SW networks has gained tremendous interests of
scientists. A variety of models have been proposed to explain the
underlying mechanisms that ensure finding shortest paths based
exclusively on local information
\cite{Kleinberg2000,Watts2002,Krioukov2008}. In these models,
networks were generated based on underlying reference frames, e.g.
grids, hierarchy, and hyperbolic spaces, which determined how networks
were organized. Vertices were contained in the underlying reference
frames which provided definitions of distances between vertices, and
adjacent vertices were more likely to be connected. Navigation was
modeled by greedy routing: messages were sent to one neighbor
nearest to the target in underlying reference frames, which was
efficient for passing messages if vertices were aware of positions of
its neighbors and targets. Indeed, the aforementioned works suggest
that networks act as an overlay on
underlying reference frames during navigation. Therefore,
navigability of networks are based on the fact that the underlying
reference frames are navigable. In these models, efficient
navigation needs prior knowledge about organization of networks.
However, several real large-size networks, e.g. email networks and online
social service networks, are self-organized, so that it's hard for
individuals to be aware of underlying reference frames and discover
their exact positions.

Here, we aim to address the navigability of SW networks through a
different way: establishing a general scheme for efficient
navigation, regardless of the underlying reference frames of networks.
This kind of problems have been considered before. One method is to
reconstruct underlying reference frames, e.g. embedding networks
generated by Kleinberg's model into Euclidean plane and
reconstructing the dimension of the underlying lattice when network
generated by long-range percolation
\cite{Sandberg2006,Benjamini2001}. The other method is to embed a
network into a metric space and ensure that distances between
vertices are proportional to shortest path length through a proper
embedding algorithm, regardless of underlying reference frames
\cite{Francis1999,Ng2002,Dabek04}.

We construct a scheme for navigation following the idea of metric
space. The embedding algorithm is inspired by the fact that
navigation on social networks is based on information exchanged and
accumulated by communication, which is used to determine who among
the acquaintances are `socially closest' to target persons. Therefore
we embed networks into a metric space through a process of information
exchange and accumulation, in which vertices find their positions by
distributed and localized self-organization. It is demonstrated by
numerical simulations that the self-organized algorithm can establish
a scheme for efficient navigation, irrespective of the underlying
reference frames of networks, and we find that the navigability of networks
is influenced by SW properties which are characterized by low diameter
and high cluster coefficient.

\section{Algorithm to Establish Navigation Scheme}

The key for addressing the navigability lies in the self-organized
embedding algorithm in the absence of prior knowledge about
underlying reference frames. In our algorithm, an $m$-dimension
Euclidean space is chosen as the metric space to define distances
between vertices. Then we follow the self-organized process of
information exchange and accumulation on social networks, which is
described as follows
\begin{equation}
\mathbf{x}_{i,t} =
f(\mathbf{x}_{j,t-1}),j\in\mathbf{N}_{i},\label{xe}
\end{equation}
\begin{equation}
\mathbf{p}_{i,t} = \mathbf{p}_{j,t-1}+\mathbf{x}_{i,t}, \label{pe}
\end{equation}
where $\mathbf{N}_{i}$ is the set of immediate neighbors of vertex
\emph{i}. Vectors $\mathbf{x}$ and $\mathbf{p}$ consist of
$m$ elements corresponding to the $m$ dimensions of metric space.
The vector $\mathbf{x}$ is coupled through the network topology and
simultaneously updated according to Eq. (\ref{xe}), while position
vector $\mathbf{p}$ is the cumulative summation of historical vector
$\mathbf{x}$. Since information exchange in Eq. (\ref{xe}) is
restricted between vertices and their direct neighbors, the
algorithm is distributed and localized. Meanwhile, distances between
vertices will be constant if vector $\mathbf{x}$ can converge after
sufficient evolving steps. Moreover, vertices can be seen as
flocking in a metric space, and vectors $\mathbf{x}$ and
$\mathbf{p}$ represent velocity and position, like in Vicseck model
\cite{Vicsek1995}. Velocities of tightly connected vertices
synchronize more quickly. Therefore, vertices connected by shorter
paths will gather in the metric space, which ensures that the
distances between vertices in the metric space are associated with
path lengths on networks. Messages can be delivered along short
paths by navigation based on distances in the metric space.

Many dynamics can be applied as a realization of Eq. (\ref{xe}),
such as chaotic oscillators coupled by networks which can
synchronize depending on suitable coupling strengths. For the
purpose of its simplicity, we choose the updating rule of vector
$\mathbf{x}$ as follows: at every time step, values of
$\mathbf{x}_i$ is the average of its neighbors and the initial
$\mathbf{p}_{i,0}$ equals to $\mathbf{x}_{i,0}$. Then the algorithm
can be written as
\begin{equation}
\mathbf{p}_{0} = \mathbf{x}_{0}, \label{P0}
\end{equation}
\begin{equation}
\mathbf{x}_{i,t} =
\frac{1}{d_i}\sum_{j}\mathbf{x}_{j,t-1},j\in\mathbf{N}_{i},\label{xa}
\end{equation}
\begin{equation}
\mathbf{p}_{i,t} = \mathbf{p}_{i,t-1}+\mathbf{x}_{i,t}, \label{Pa}
\end{equation}
where $d_i$ is the degree of vertex $i$. Equations (\ref{xa}) and
(\ref{Pa}) can be rewritten in matrix form as the combinations of
eigenvectors of normal matrix $\mathbf{N}$ of network
\begin{equation}
\mathbf{P}_{0} = \mathbf{X}_{0} = \mathbf{V}\mathbf{A}, \label{X0c}
\end{equation}
\begin{equation}
\mathbf{X}_{t} =
\mathbf{N}\mathbf{X}_{t-1}=\mathbf{N}^t\mathbf{X}_{0}=\mathbf{V}\mathbf{D}^t\mathbf{A},
\label{Xc}
\end{equation}
\begin{equation}
\mathbf{P}_{t} =
\mathbf{P}_{t-1}+\mathbf{X}_{t}=\mathbf{V}(\mathbf{I}+\sum_{i=1}^{t}\mathbf{D}^i)\mathbf{A}.\label{Pc}
\end{equation}
Every row of matrices $\mathbf{X}$ and $\mathbf{P}$ is the vector of
velocities and positions of each vertex. Columns of matrix
$\mathbf{V}$ are the eigenvectors of $\mathbf{N}$. $\mathbf{A}$
consists of linear combination coefficients when eigenvectors of
$\mathbf{N}$ are chosen as basis vectors. Matrix $\mathbf{D}$ is a
diagonal matrix with eigenvalues of normal matrix on the main
diagonal. Because eigenvalues of $\mathbf{N}$ are in the
interval $[-1,1]$, for long enough evolving time $t$, we get the
final position matrix $\mathbf{\tilde{P}}$ as
\begin{equation}
\mathbf{\tilde{P}}=\mathbf{V}\mathbf{E}\mathbf{A}. \label{Pfinal}
\end{equation}
Matrix $\mathbf{E}$ is a diagonal matrix whose $i$th diagonal
element is $1/(1-\lambda_i)$, where $\lambda_i$ is the eigenvalue of
normal matrix. It can be seen that eigenvectors corresponding to large
eigenvalues play more important roles in the position matrix as a result
of the factor $1/(1-\lambda_i)$.

Since the positions of vertices in the metric space are linear combinations
of eigenvectors of normal matrix, it demonstrates that the
embedding can represent network topology, which is reflected by the
fact that adjacent vertices in the metric space are connected by shorter
paths on network. The distance between vertex $i$ and $j$ after
sufficient evolving time is
\begin{equation}
\begin{split}
&d_{i,j}^2=\sum_{l=1}^{m}\left[\sum_{k=1}^{n}\frac{a_{k,l}}{1-\lambda_k}(v_{i,k}-v_{j,k})\right]^2\\
&=\sum_{l=1}^{m}\sum_{k=1}^{n}\left[\frac{a_{k,l}}{1-\lambda_k}\right]^2((v_{i,k}-v_{j,k}))^2+\\
&2\sum_{l=1}^{m}\sum_{p=1}^{n}\sum_{q=p+1}^{n}\frac{a_{p,l}a_{q,l}}{(1-\lambda_p)(1-\lambda_q)}(v_{i,p}-v_{j,p})(v_{i,q}-v_{j,q}),
\label{d1}
\end{split}
\end{equation}
where $a_{i,j}$ and $v_{i,j}$ are the elements of matrix
$\mathbf{A}$ and $\mathbf{V}$, respectively. If elements of
$\mathbf{X}_0$ are uniformly distributed in interval $[-1 ,1]$, the
elements of matrix $\mathbf{A}$ have the following properties:
$\langle a_{i,j}\rangle=0, \langle a_{i,j}a_{k,l}\rangle=0$ and
$\langle a_{i,j}^2\rangle=\langle x^2\rangle$. In addition, if $m$
is sufficiently large, the distance can be expressed by
\begin{equation}
d_{i,j}^2=\sum_{k=1}^{n}\frac{m\langle
x^2\rangle}{(1-\lambda_k)^2}(v_{i,k}-v_{j,k})^2. \label{d3}
\end{equation}
Equation (\ref{d3}) shows that the distances between vertices can
be seen as those in the situation that position values of vertices
are elements of weighted eigenvectors of normal matrix. Due to the
factor $(1-\lambda_k)^{-2}$, distances are mostly determined by
eigenvectors associated with large eigenvalues. It has been
proved that these eigenvectors are the solutions of following
constrained optimization problem \cite{Capocci2004}. Let the energy
of system $z(x)$ be defined as
\begin{equation}
z(x)=\frac{1}{2}\mathbf{x'Lx}, \label{z}
\end{equation}
where $\mathbf{L}$ is the Laplace matrix of network and $\mathbf{x}$
are position values assigned to the vertices together with a
constraint
\begin{equation}
\mathbf{x'Kx}=1, \label{constraint}
\end{equation}
where matrix $\mathbf{K}$ is a diagonal matrix whose $i$th main diagonal
element is the degree of vertex $i$. Let
$\lambda_1 < \lambda_2 < \ldots < \lambda_{n-1} < \lambda_n=1$ be the
eigenvalues, and the corresponding eigenvectors under constraint of
Eq. (\ref{constraint}) are $v_1, v_2 , \ldots, v_{n-1}$ and $v_n$. The
minimum nontrivial value of $z$ is $1-\lambda_{n-1}$, and the
relevant position vector $x$ is $v_{n-1}$. If the energy reaches
the minimum nontrivial value, vertices which are connected by a
number of short paths are sufficiently close in the metric space
constructed by eigenvectors, which ensures that distances in the metric
space correspond to path lengths on networks.

Due to the fact that similar vertices are more likely to be connected,
it's natural to evaluate similarities based on the number of paths
between vertices and the length of paths in the absence of prior
knowledge of underlying reference frames \cite{Leicht2006}. Through this
evaluation, vertices connected by more and shorter paths, which will
be adjacent in the metric space after self-organized embedding, are
deemed to be more similar. Therefore, the results of embedding
algorithm are in consistent with the basic ideas of underlying reference frames:
similar vertices are adjacent, and more likely to be connected.

\section{Experimental Results}

\subsection{Experimental Results of Small-World Networks Generated by
WS Model}

\begin{figure*}[!t]
\centering \subfigure[\ diameter and cluster
coefficient]{\label{fig:pddcc}\includegraphics[width=3.3in]{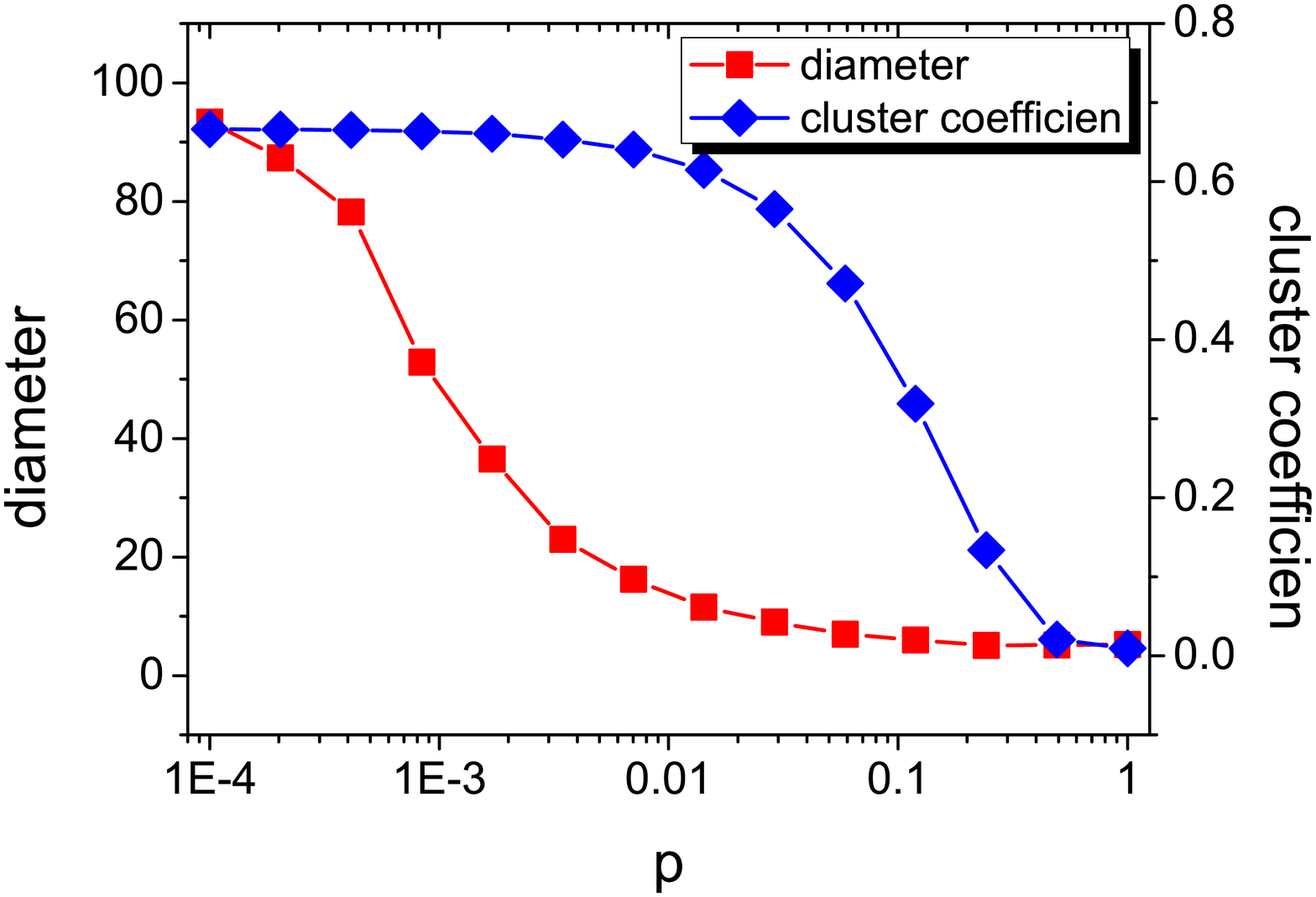}}
\hspace{0.3in} \subfigure[\
$m=5$]{\label{fig:psucstr5}\includegraphics[width=3.3in]{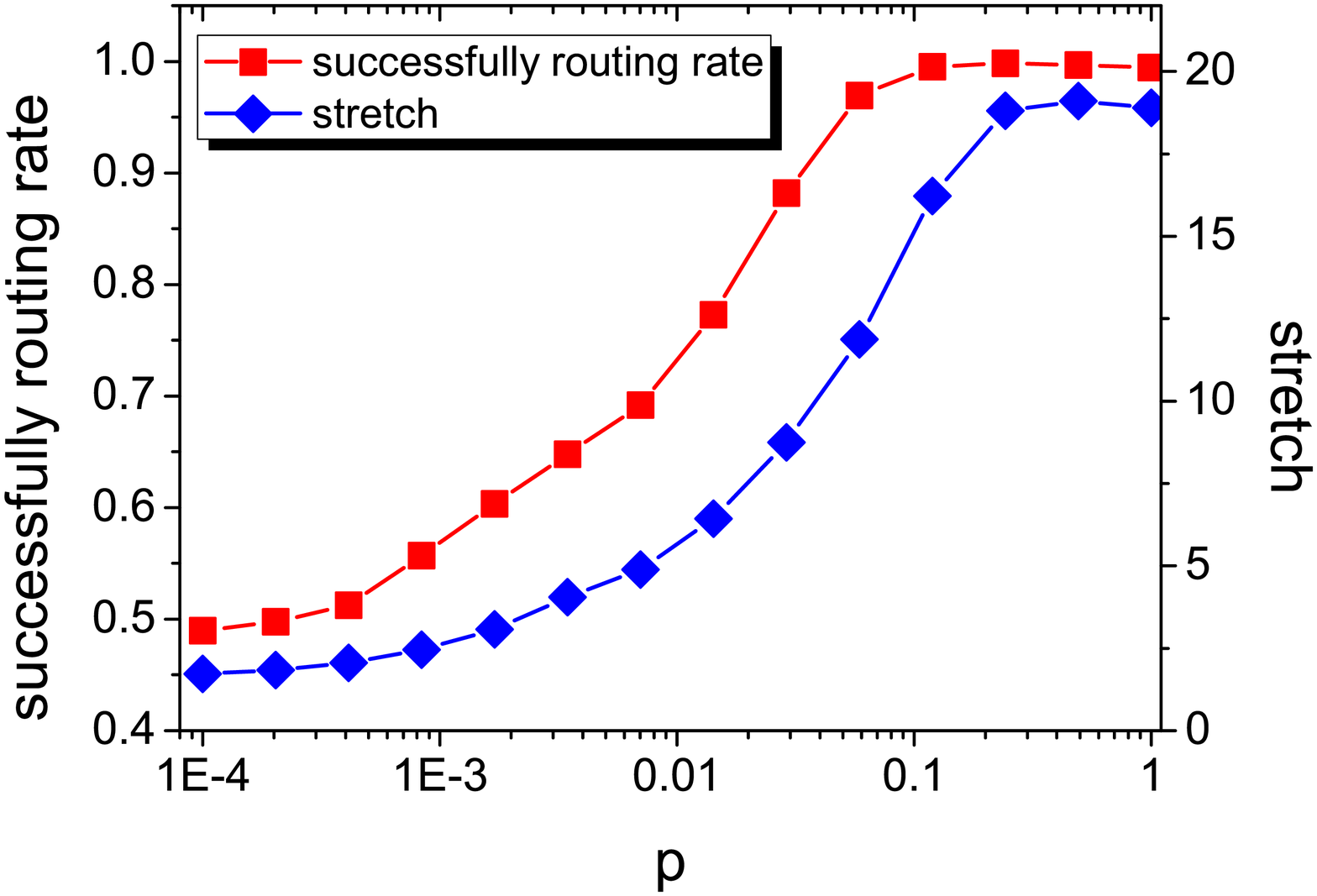}}
\\
\subfigure[\
$m=10$]{\label{fig:psucstr10}\includegraphics[width=3.3in]{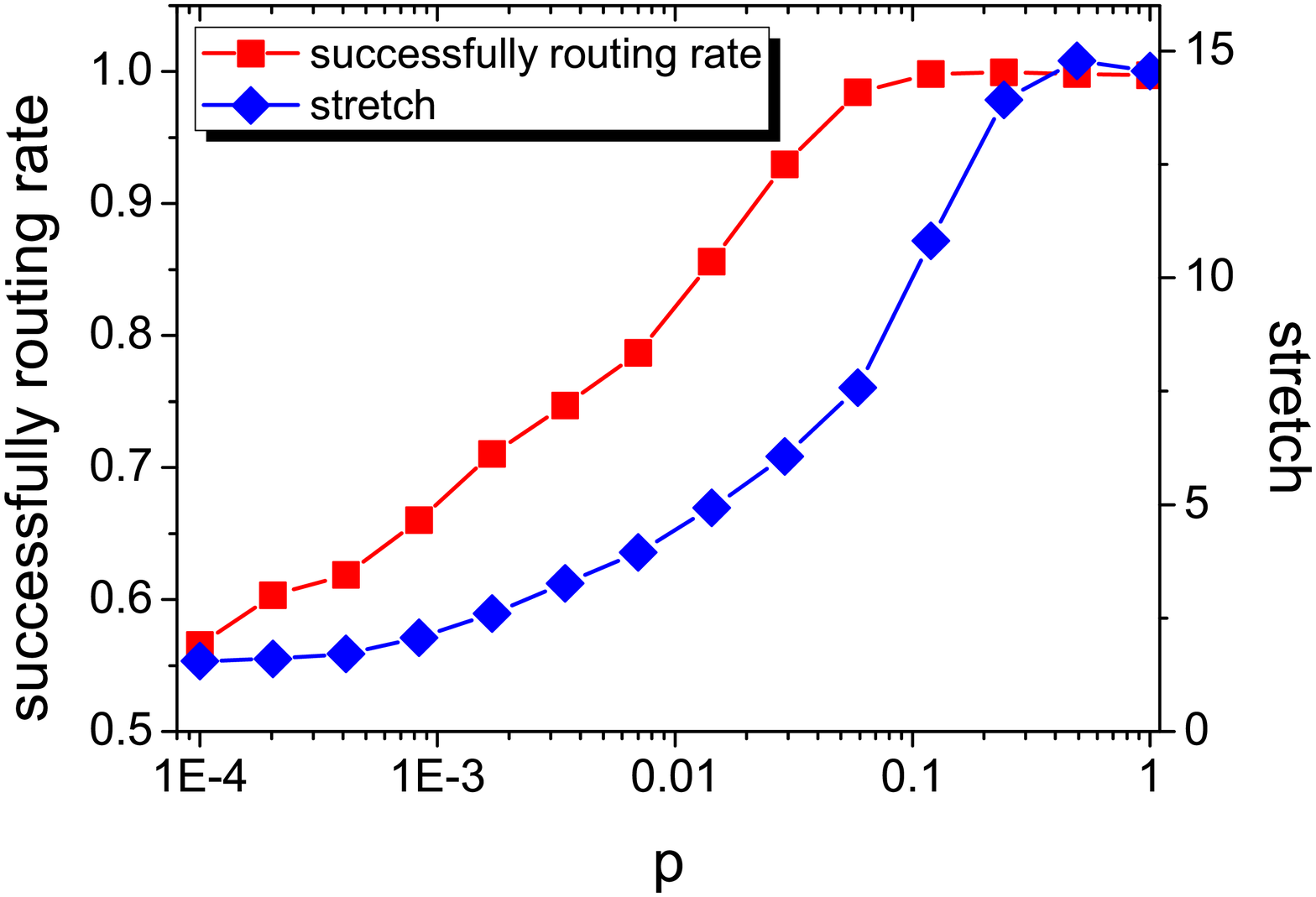}}
\hspace{0.3in} \subfigure[\
$m=20$]{\label{fig:psucstr20}\includegraphics[width=3.3in]{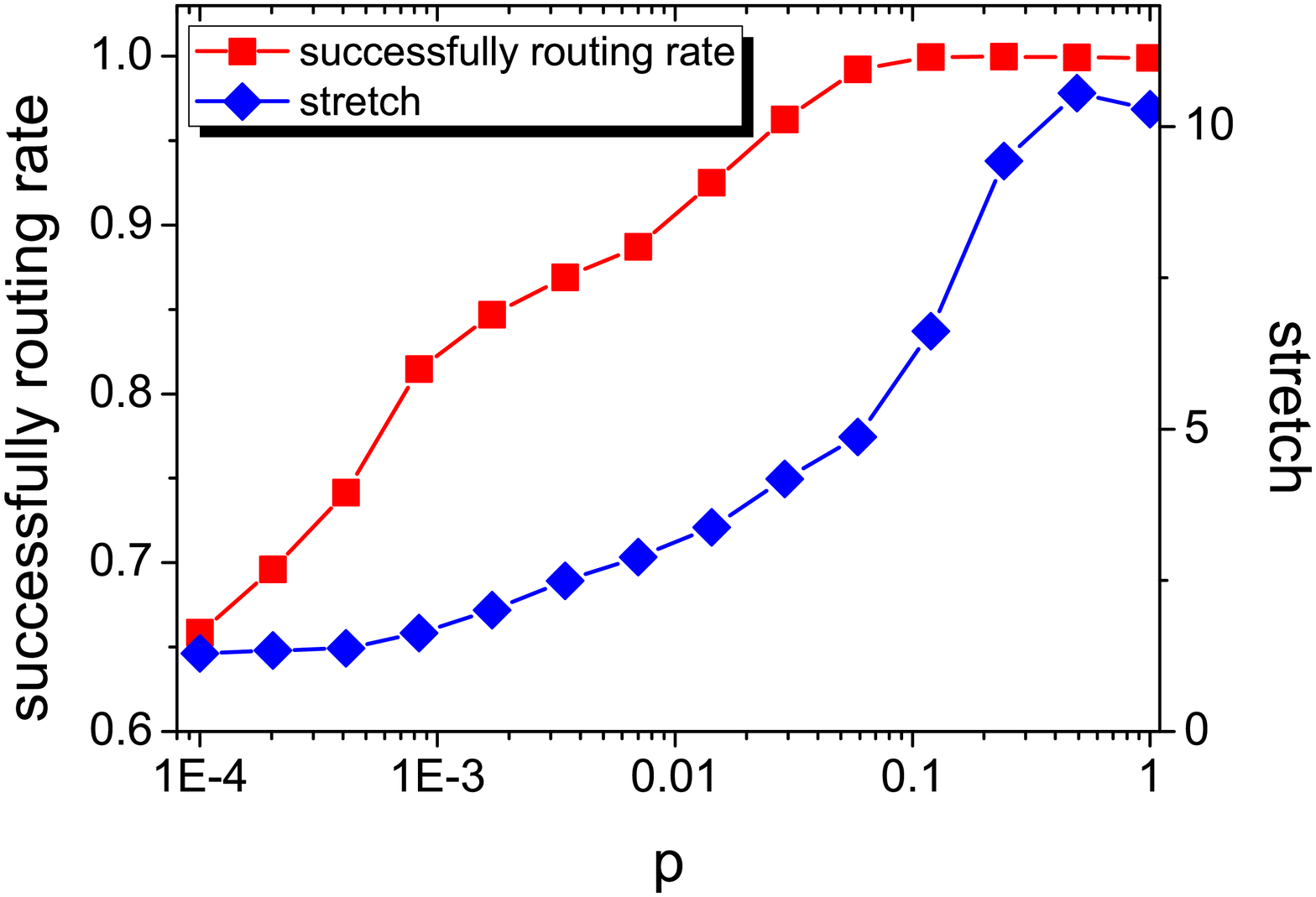}}
\caption{(Color online) Diameter and cluster coefficient as a
function of rewiring probability $p$ (a); performance of greedy routing
for different dimension of metric space (b) $m=5$, (c) $m=10$ and
(d) $m=20$. Networks are generated by WS model \cite{Watts1998}.
Numerical simulations at each $p$ are averaged over $20$ realizations of model. SW
networks show strong navigability with high successfully routing
rate and low stretch for all dimensions. In particular, the SW
properties are necessary for navigability, and the large metric space
dimension is useful to improve navigability.} \label{fig:swnav}
\end{figure*}

The self-organized embedding algorithm is applied to build navigation scheme on
SW networks generated by WS (Watts-Strogatz) model, in which SW
properties result from rewiring edges of original regular network at
probability $p$ \cite{Watts1998}. The chosen original
regular network has $n$=1000 vertices, and each vertex link to $k$=10
nearest others. The diameters and cluster coefficients of networks
at different rewiring probabilities are shown in Fig. \ref{fig:pddcc}.
Experimental results are averaged over $20$
network realizations. As shown in Fig \ref{fig:pddcc},
even for the small rewiring probability, the
diameters of networks decrease sharply while the cluster
coefficients are nearly the same as the original regular
network.

At the beginning of embedding algorithm, every vertex is assigned an
initial velocity $\mathbf{x}_{i,0}$, whose values of each dimension
are uniformly distributed within $[-0.5,0.5]$. Dimensions of metric
spaces chosen to be $m=5$, $10$ and $20$ are to investigate how the
metric space influences navigation. The embedding algorithm is
terminated when the velocities of vertices reaches a certain
synchronization level. We defined the synchronization error of
$\mathbf{x}_i$ of dimension $k$ at evolving time $t$ as
\begin{equation}
e_t(k)=\frac{1}{n}\sum_{i=1}^n(x_{i,t}(k)-\langle
x_{i,t}(k)\rangle)^2.\label{er}
\end{equation}
When the synchronization errors of velocities at each dimension are
less than a small value, which is chosen as $10^{-4}$, the
embedding algorithm is terminated.

The greedy routing strategy to simulate navigation on networks can
be described as follows: vertices are aware of positions of their
neighbors in the metric space and positions of targets are transmitted by
messages. Messages are passed through current hop to the neighbor closest to
targets at each step. To avoid loops, messages are prohibited from
neighbors that have been visited. The routing will terminate if
message reaches target or all the neighbors of current hop have been
visited. We randomly pick $10^4$ source and target pairs for every
network to be navigated. Notice that the navigation is not
symmetric, e.g. navigation from vertex $i$ to $j$ is not equivalent
to navigation from $j$ to $i$ because the local environments of
vertex $i$ and $j$ are different. Efficient navigation is defined by the fact that
messages are successfully passed to targets along the shortest
paths. Therefore, we examine two metrics to evaluate navigability:
the successfully routing rate (the ratio of number of successfully
routing messages and all messages) and the stretch (average of the
ratios of routing path length and shortest path length of each
message).

Figure 1(b), (c) and (d) show that the successfully routing
rates and stretches are as a function of rewiring probability $p$ for the hidden
metric space of different dimension. When rewired connections start
to arise, successfully routing rates increase quickly, whereas
stretches grow much slower until cluster coefficients drop sharply.
As a result of the different growing speeds, high successfully routing
rates and low stretches, which indicates the efficient
navigation and strong navigability, simultaneously occur when
the networks show small-world properties, and are much more apparent for the
hidden metric space of larger dimensions. In other words, the larger dimension
of hidden metric space is useful to improve performances of
navigation, which is reflected by higher successfully routing
rates and lower stretches at the same rewiring probability.

\begin{figure*}[!t]
\centering \subfigure[$\
p=0.0001$]{\label{fig:p0001}\includegraphics[width=3.3in]{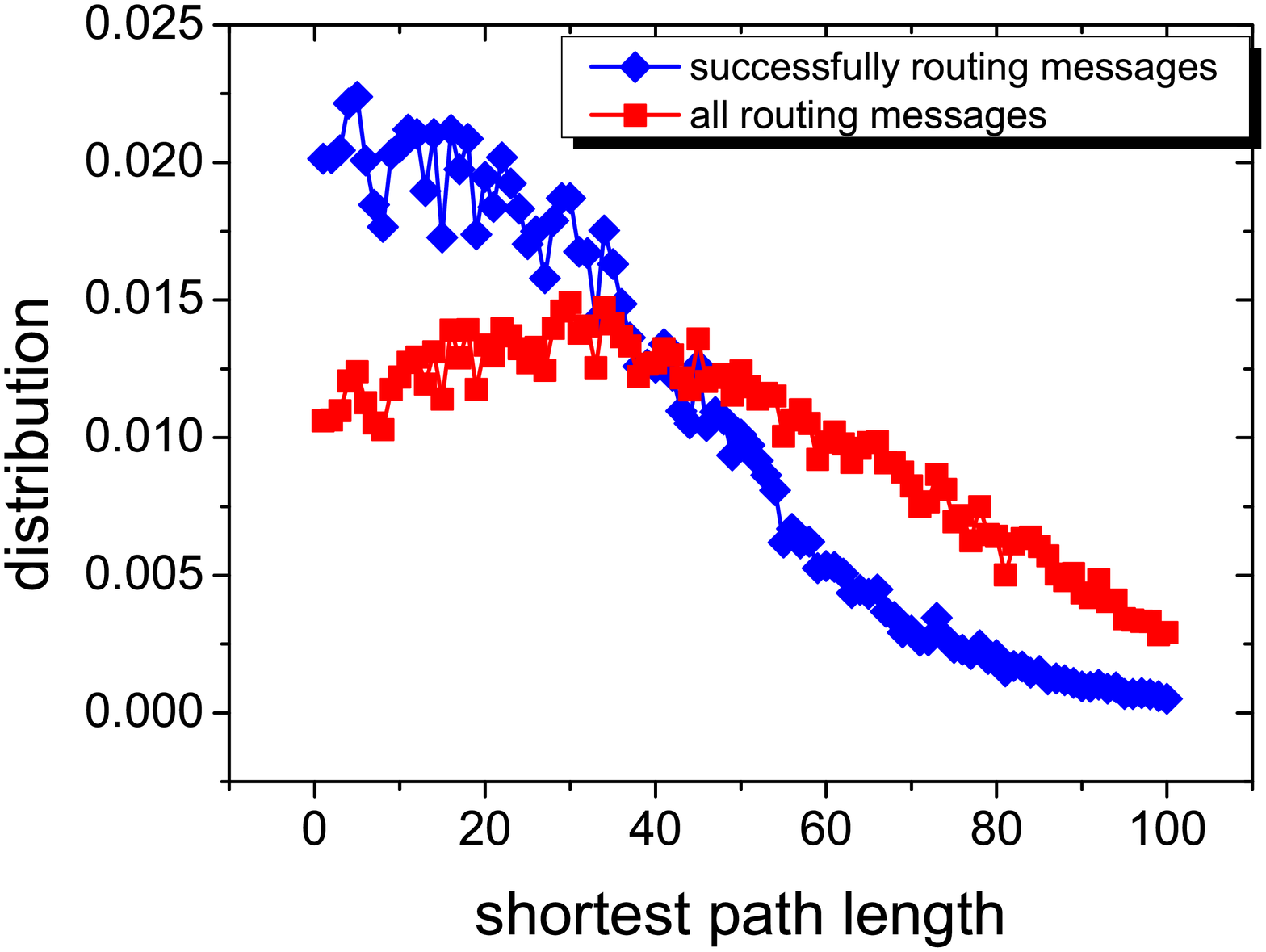}}
\hspace{0.3in} \subfigure[$\
p=0.0002$]{\label{fig:p0002}\includegraphics[width=3.3in]{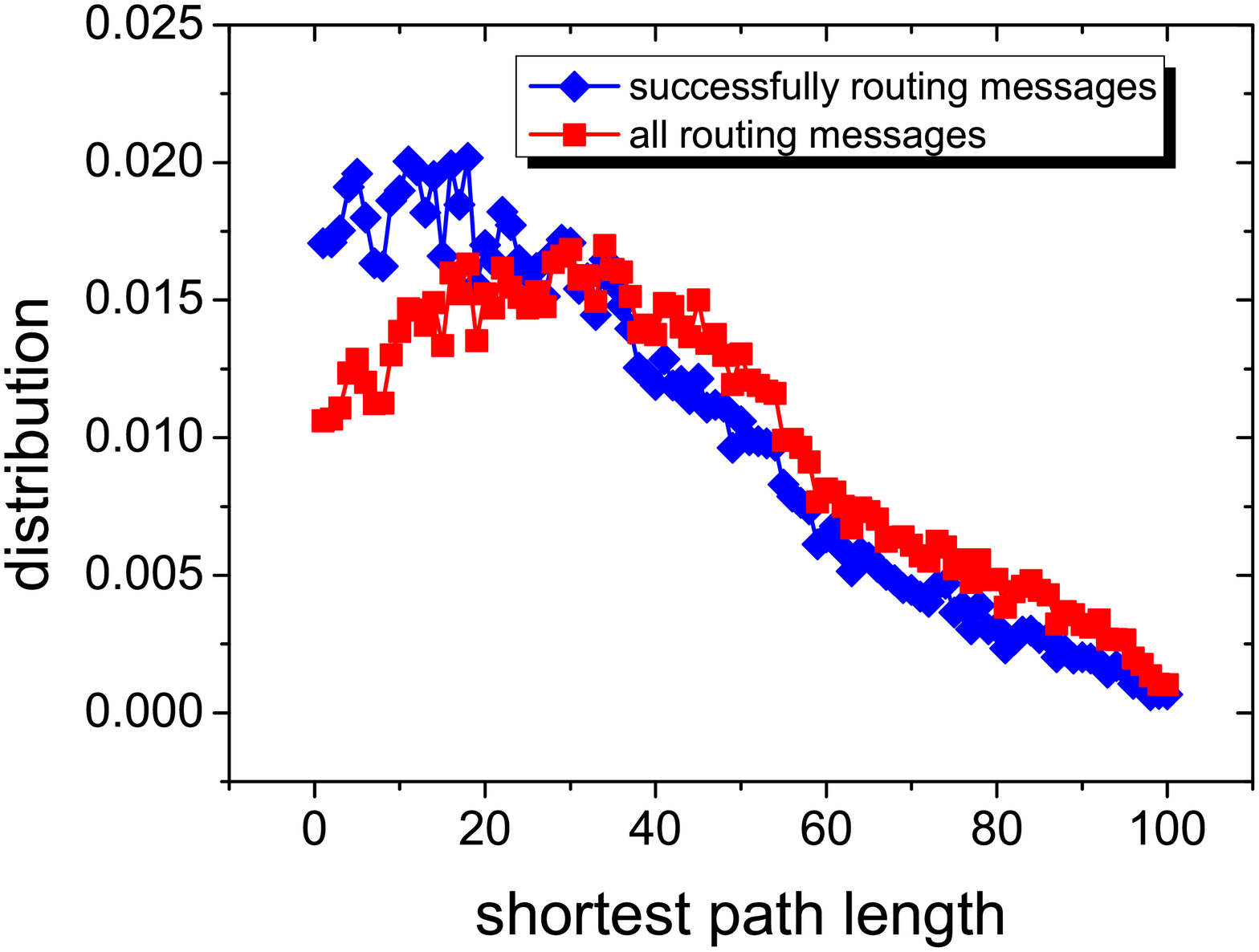}}
\\
\subfigure[$\
p=0.0004$]{\label{fig:p0004}\includegraphics[width=3.3in]{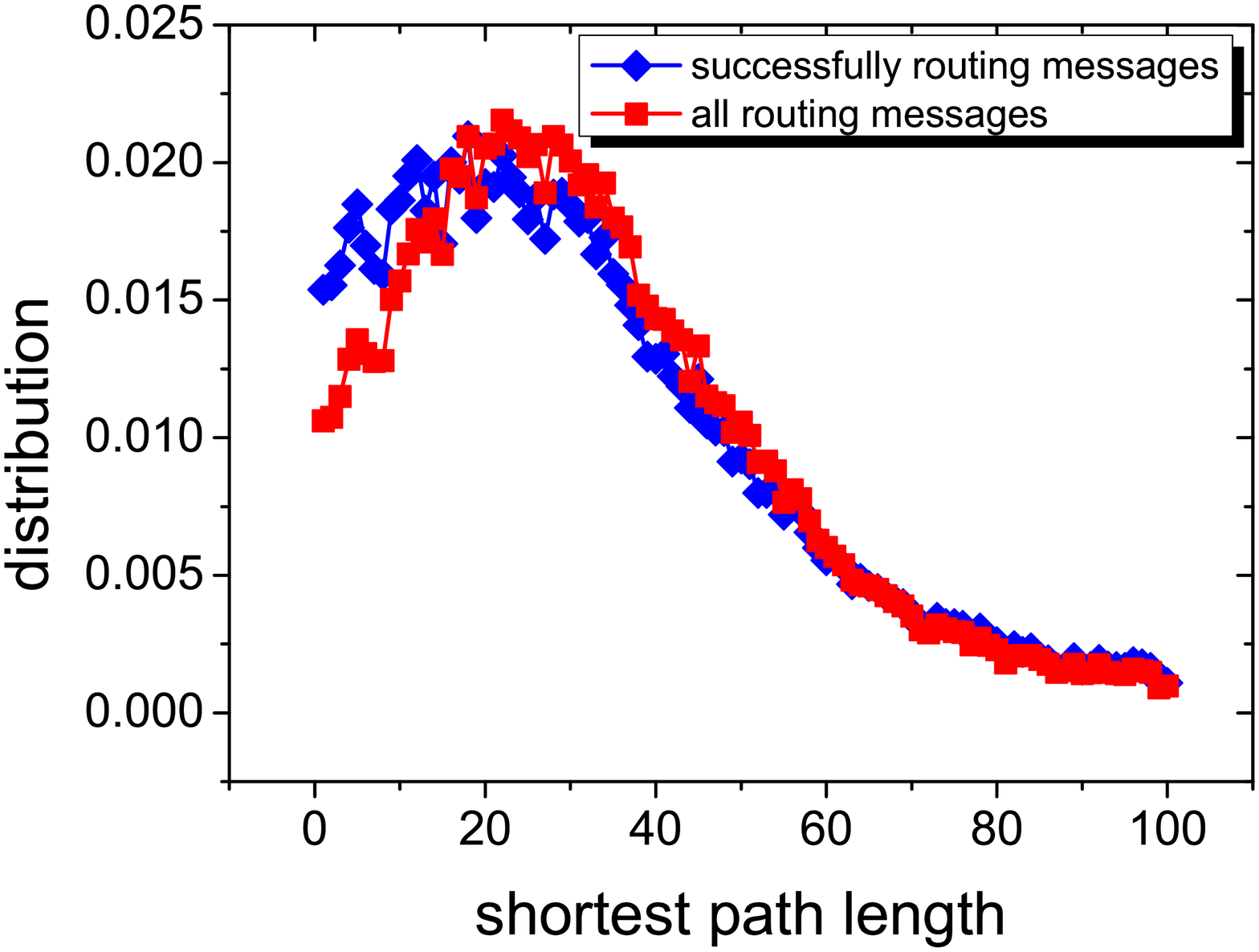}}
\hspace{0.3in} \subfigure[$\
p=0.0008$]{\label{fig:p0008}\includegraphics[width=3.3in]{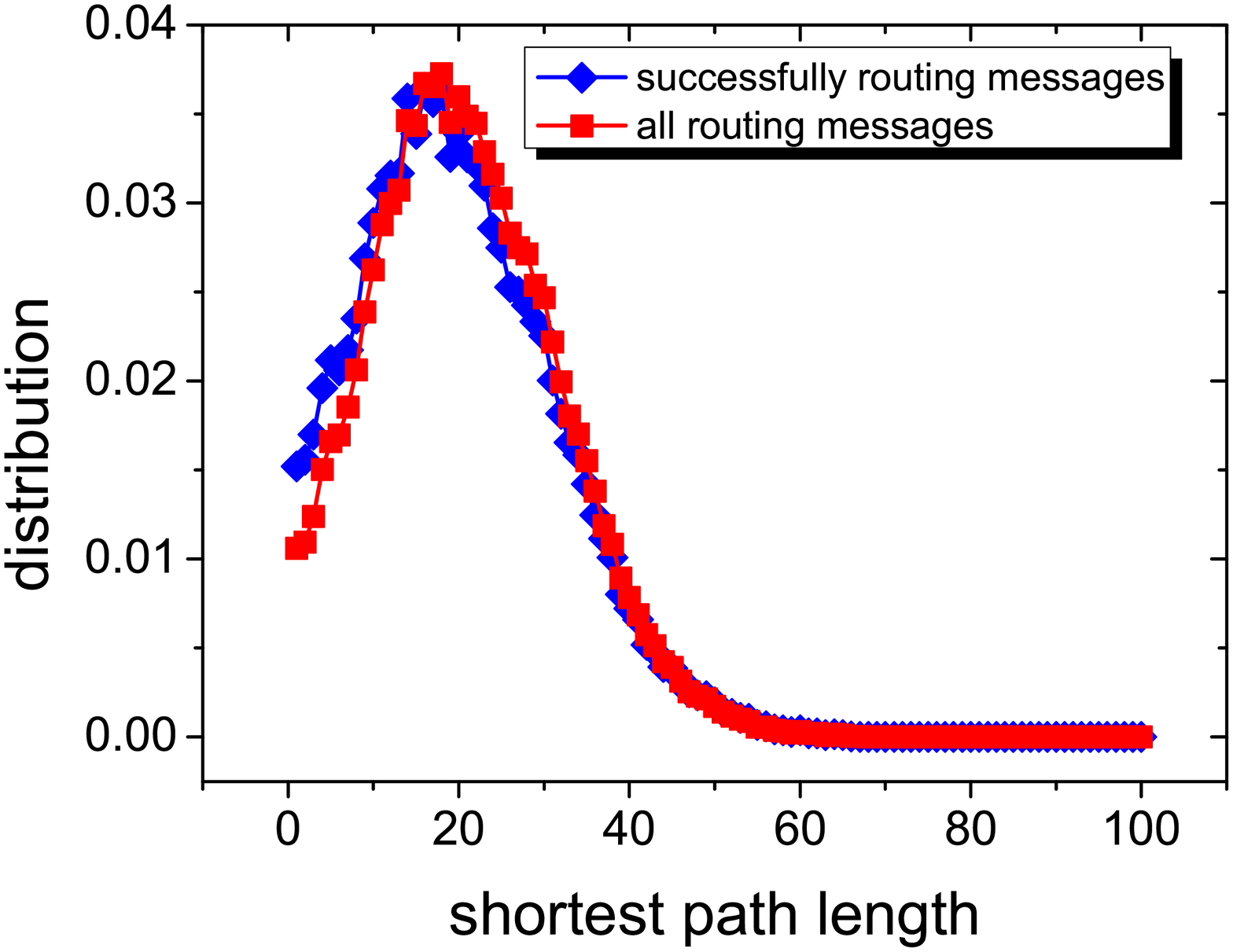}}
\caption{ (Color online) Distributions of shortest path length
between vertices of all routing messages and successfully routing
messages for different rewiring probability $p$:(a) $p=0.0001$, (b)
$p=0.0002$, (c) $p=0.0004$ and (d) $p=0.0008$. As the number of long range
connections increases, messages can escape from the local area of
source vertices and travel a long distance. The two distribution
have agreed well with each other even when the rewiring probability is
still very small. These results reflects the power of weak ties: even
if a few long range connections exist, messages could be passed to
the whole network.} \label{fig:longrange}
\end{figure*}

Long range connections, or the so-called weak ties in sociology,
play an important role in activities on networks, e.g. information
which people receive through weak ties is more useful and
successfully routing messages on Email networks are conducted
primarily through intermediate to weak strength ties
\cite{Granovetter1973,Dotts2003}. Hence it is worthy of studying
that long range connections affect navigation by passing messages to
vertices far away from each other on networks. We calculate the
distributions of shortest path length between all pairs and
successfully routed pairs at different rewiring probability $p$ when
the metric space dimension is 20 (see in Fig. \ref{fig:longrange}).
When there are fewer long range connections, navigation finds more
nearby targets than those far away. The reason is that many messages
can't travel far away from the start vertices on networks with
high cluster coefficients resulting from quickly arriving at the
vertices whose neighbors have been all visited. As the number of
long range connections increases, messages can escape from the local
area of source vertices and travel a long distance on networks to
arrive at targets. Therefore, targets are successfully reached at
the same probability for different path length from sources.
Moreover, the two distributions have agreed well with each other
when the rewiring probability is 0.0008, which results in
sufficiently small number of long-range connections compared to the
total number of connections. The weak ties are extremely useful in
the sense that even if a few long range connections exist, messages
could be passed to the whole network. This fact also explains why
the successfully routing rates immediately increase fast when there
are only a few long range connections.

Navigability of networks in terms of self-organized embedding
algorithm is based on the fact that distances in the metric space
are associated with similarities of vertices extracted from
topology. However, we can't ensure that distance between every
vertex pairs represents its similarity in the absence of central
control, e.g. adjacent vertices in the metric space may not be
tightly connected. Actually, greedy routings performed on networks
consist of two parts: properly directed part is more relevant
to navigation, while the remaining part is more tendency to random
walks. For numerical simulations on SW networks, when cluster
coefficients stay at a high value, there are clusters consisting of
tightly connected similar vertices, which satisfies the organizing
rules of networks based on underlying reference frames. In this
case, network topology can be mapped into a hidden metric space by
self-organized embedding algorithm. Meanwhile, messages cannot
travel along paths through random walk on networks with high cluster
coefficients because they are easy to reach a vertex, all of whose
neighbors have been visited. Successfully passed messages on highly
clustered networks are mostly routed by navigation, which leads to
low stretches. When cluster coefficients drops quickly, the local
clusters vanish by randomly rewired connections and vertices are
randomly placed in the hidden metric space, which differs the
embedding of networks from the network topology. In this regard,
random walks can travel a long path to reach targets because
vertices have little common neighbors. Therefore, most messages are
successfully delivered by random walks, which leads to large
stretches and the successfully routing rates are close to 1.

\subsection{Experimental Results of Small-World Networks with Power-Law Degree Distribution}

\begin{figure*}[!t]
\centering \subfigure[\ diameter and cluster
coefficient]{\label{fig:rddcc}\includegraphics[width=3.3in]{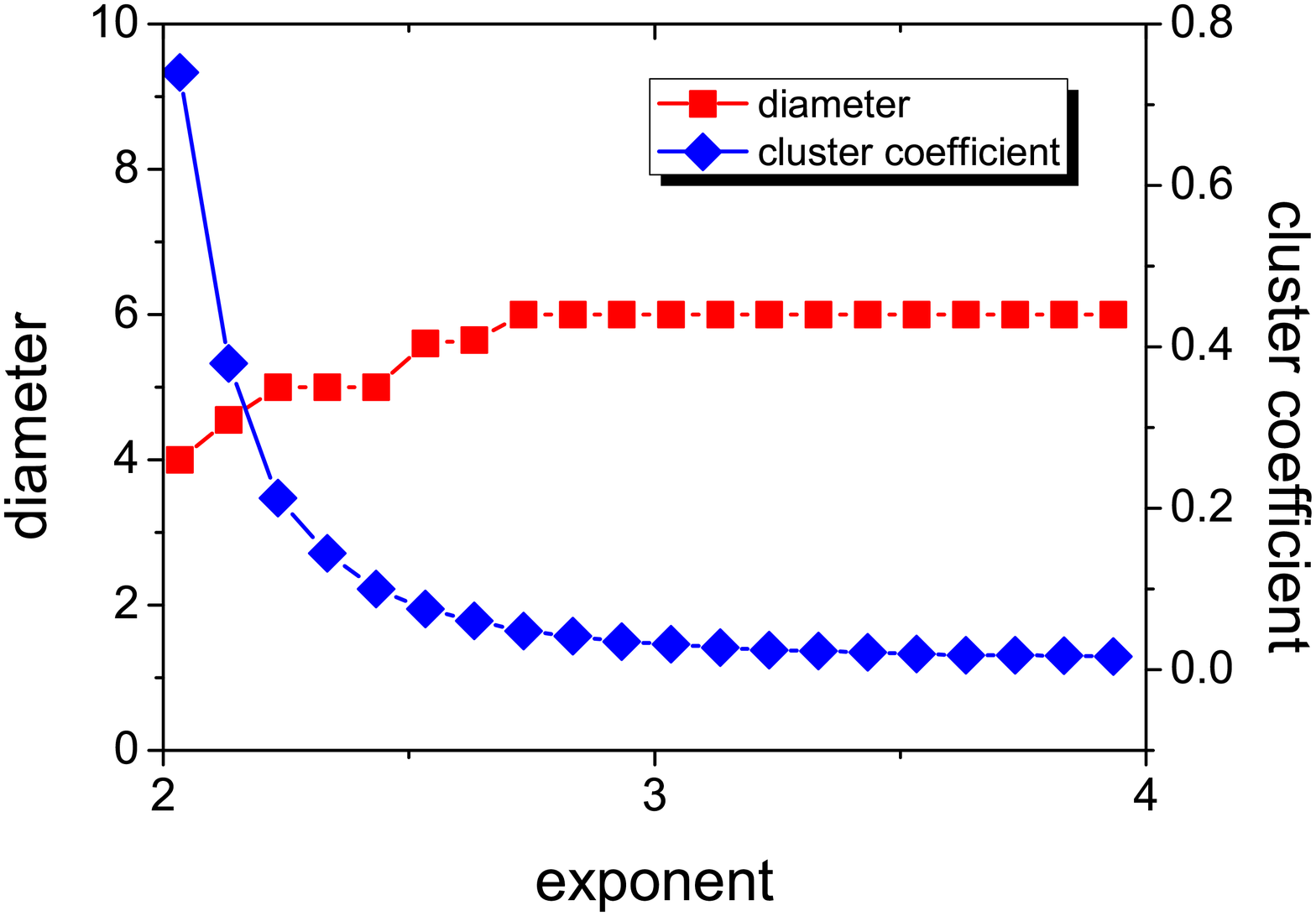}}
\hspace{0.3in} \subfigure[\
$m=5$]{\label{fig:rsucstr5}\includegraphics[width=3.3in]{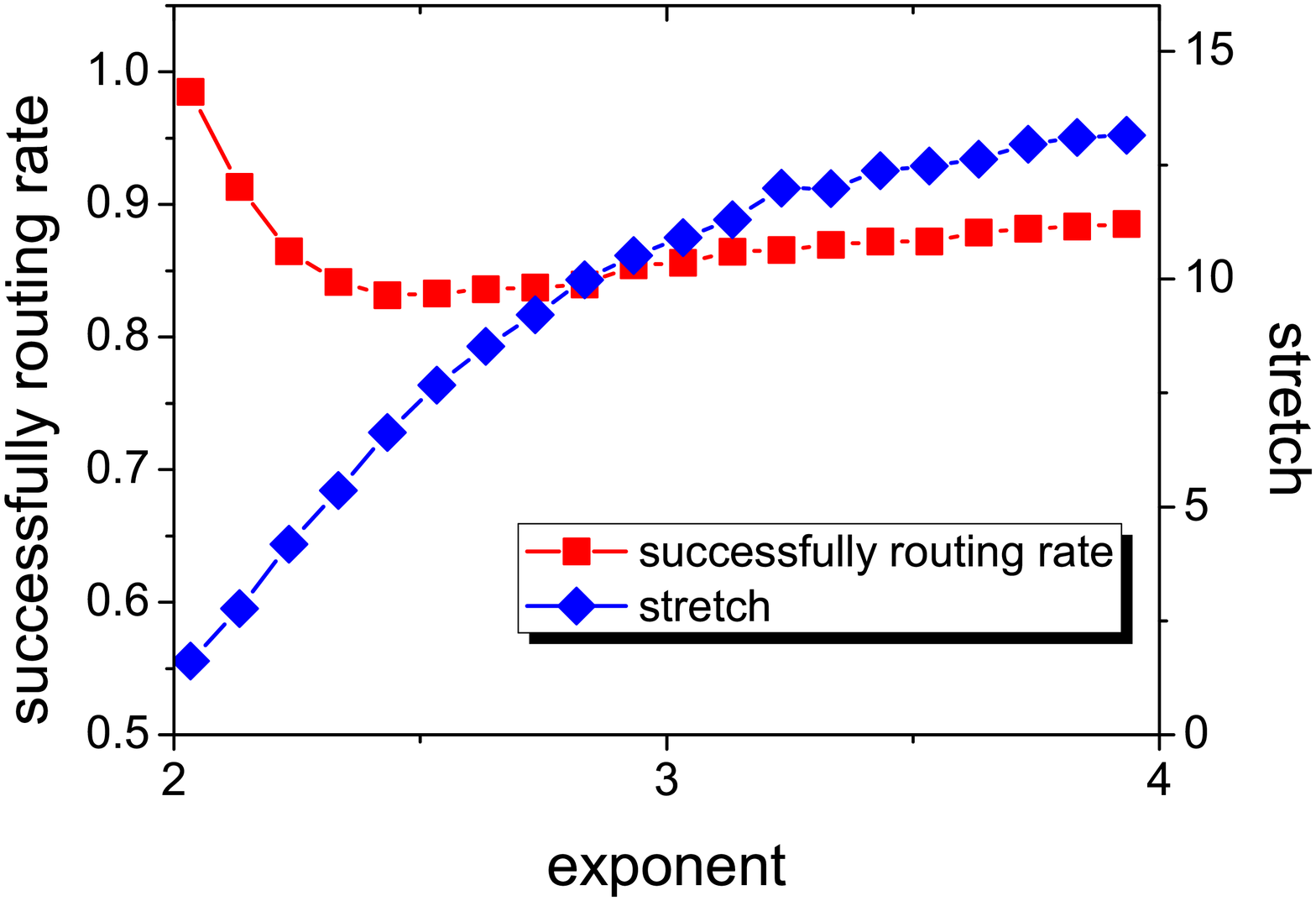}}
\\
\subfigure[\
$m=10$]{\label{fig:rsucstr10}\includegraphics[width=3.3in]{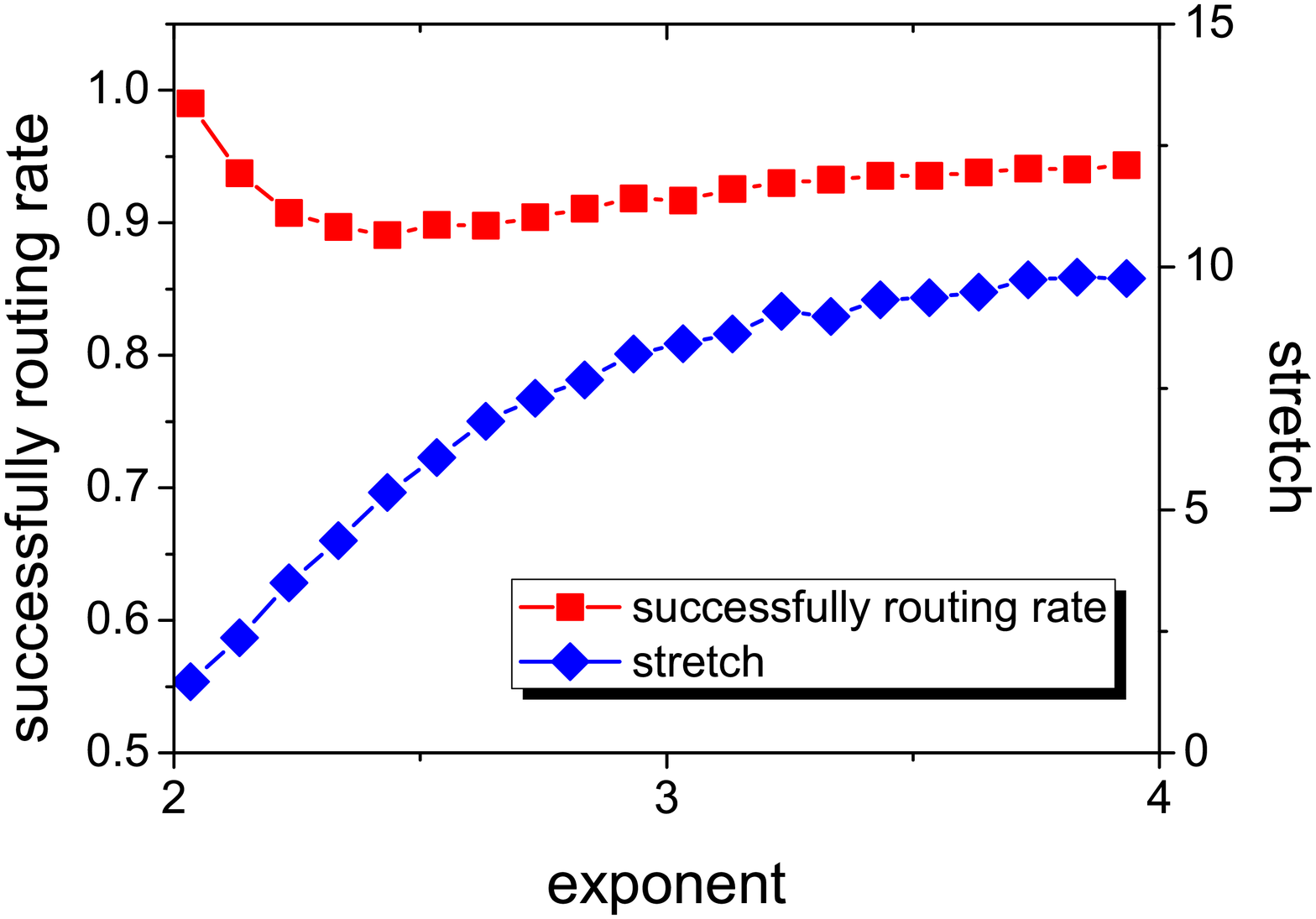}}
\hspace{0.3in} \subfigure[\
$m=20$]{\label{fig:rsucstr20}\includegraphics[width=3.3in]{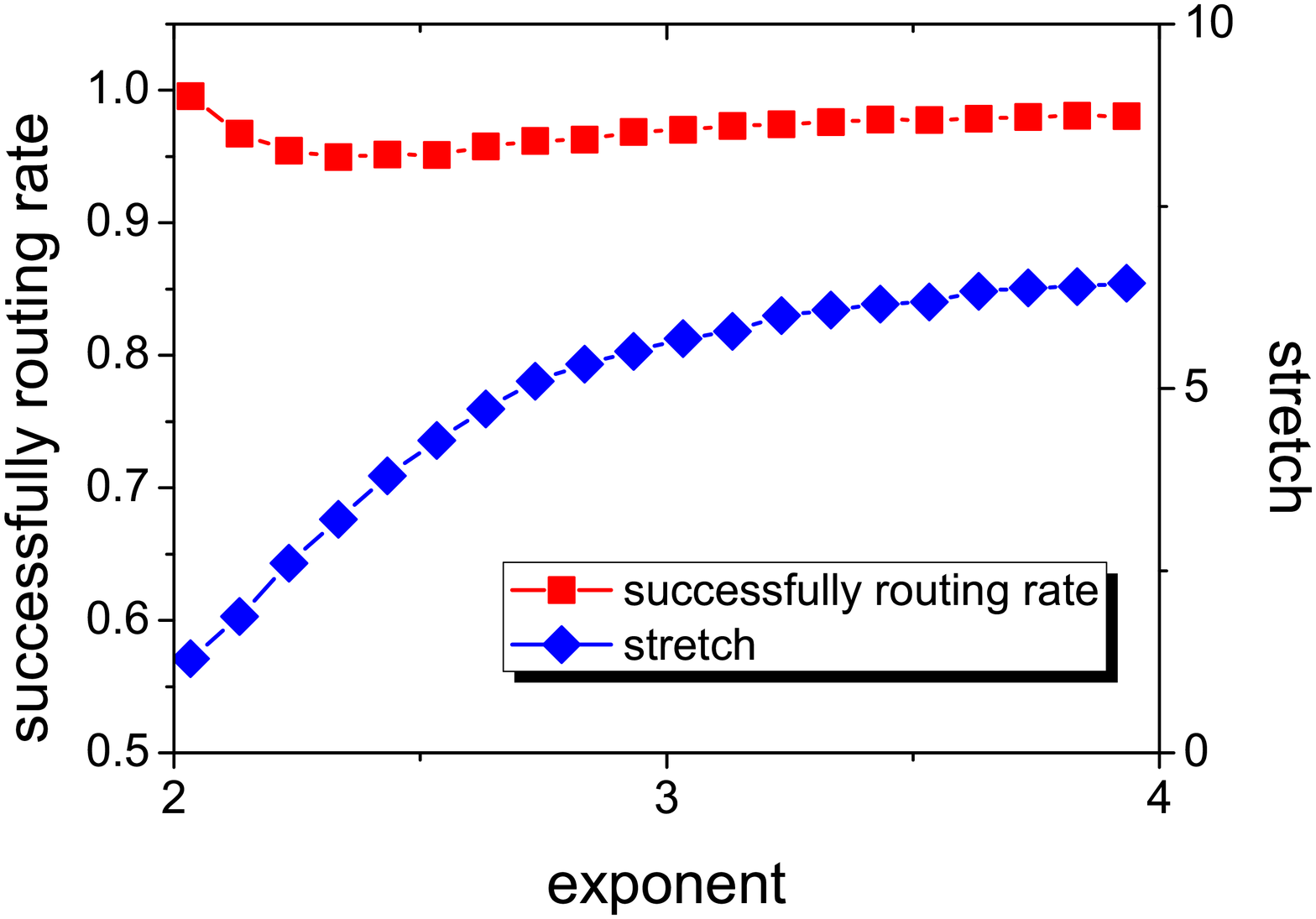}}
\caption{ (Color online) Diameter and cluster coefficient as a
function of exponent $\gamma$ (a); performance of greedy routing for
different dimension of metric space (b) $m=5$, (c) $m=10$ and (d)
$m=20$. Networks are generated by the generalized BA model
\cite{Dorogovtsev2000,Krapivsky2001}. Experimental results at each $\gamma$ are
averaged over $20$ realizations of the model. Scale-free networks
with small exponent $\gamma$ show strong navigability represented by
high successfully routing rate and low stretch for all dimensions.
Like the results of SW models, both high cluster coefficient and low
diameter are also necessary for navigability, and the large metric space
dimension is also helpful to improve navigability.}
\label{fig:sfnav}
\end{figure*}

Many real SW networks have the power-law degree distribution
$p(k)\sim k^{-\gamma}$, such as the Internet and WWW. They are
called scale-free networks in which there are vertices with much
larger degrees than randomly connected networks, such as ER
(Erd\"os-R\'enyi) model. The largest degree of scale-free network is
proportional to $N^{1/(\gamma -1)}$, where $N$ is the number of
vertices in networks. The BA (Barab\'asi-Albert) model has been
proposed to explain the emergence of power-law degree distributions
based on the ideal of preferential attachment \cite{Barabasi1999}.
We also investigate the navigability of scale-free networks
generated by the generalized BA model
\cite{Dorogovtsev2000,Krapivsky2001}. In this model, a vertex is
added in the network with $m$ connections at each step. The
probability of attaching to an existing vertex of degree $k$ is
proportional to $k + k_0$, where the offset $k_0$ is a constant.
Note that $k_0$ being larger than $-m$ is to ensure positive
probabilities. This model yields a power-law degree distribution
with exponent $\gamma =3+ k_0/m$. Negative values of $k_0$ lead to
exponent less than $3$, which has been observed in many real complex
networks.

The scale-free networks consist of $10^3$ vertices together with
$m=3$ and the offset $k_0$ being turnable to get exponent $\gamma$
from $2$ to $4$. Results for different $\gamma$ are averaged over 20
networks. Small-world properties of scale-free networks are shown in
Fig. \ref{fig:rddcc}. It is interesting that the cluster
coefficients of scale-free networks quickly decrease with the growth
of exponent, meanwhile the diameters only increase a little. Thus,
the scale-free networks with small exponents exhibit strong SW
properties. We construct a metric space and exactly  execute greedy
routing similar to SW models. Figure 3(b), (c) and (d) show the
performance of navigation for different exponent $\gamma$. It
demonstrates that when networks exhibit SW properties with small
$\gamma$, strong navigability emerges. Stretches are also affected
by cluster coefficients because topology of highly clustered
networks can be more properly mapped into a metric space. In
addition, it can be seen that the high degree nodes act as hubs in
navigation on scale-free networks \cite{Krioukov2008}. Therefore, as
$\gamma$ increases, successfully routing rates slightly drop because
the highest degrees decrease. However, when cluster coefficients
continuously decrease, successfully routing rates start to increase
because most messages are passed by random walks, which also lead
to large stretches.

\section{Conclusion}

In conclusion, we have well investigated the self-organized
emergence of navigability on SW networks via mapping network
topology into an Euclidean hidden metric spaces through a simple
embedding algorithm inspired by information exchanging and
accumulating and established in the absence of prior knowledge of
underlying reference frames of networks. It has been demonstrated
that high navigability emerges only if networks exhibit strong
small-world properties. Because of the lacking of prior knowledge of
underlying reference frames, the self-organized embedding algorithm
can establish navigable scheme for different kinds of SW networks,
which is supported by the results of SW networks generated by WS
model and BA model.

Underlying reference frames, in which similar vertices are adjacent
and connected at higher probability, explain how real
complex networks are organized based on similarities
between individuals. Since the clustering tendency of small-world
networks satisfies the preferential attachment in underlying
reference frames, the hidden metric space based on vertices similarities
can be established by a universal algorithm, regardless of the explicit
organizing pattern of networks.

The self-organized navigation may be a possible approach available
for scalable routing on the Internet, which has gained lots
interests recently. Many algorithms have been proposed to reduce the
storage space of routing table without remarkable increase of
routing path lengths, e.g. the compact routing schemes
\cite{Abraham2004,Brady2006,Thorup2001}. The size of routing table
could be reduced to polylogarithmic of the network size in compact
routing with stretch smaller than 3, yet global topology and central
control required to build routing scheme in these algorithm have to
demand large amount of communications on networks. In our work,
since the constructing hidden metric space and greedy routing are
distributed and localized in
a self-organized way, communication are restricted between immediate
connected vertices. Meanwhile, the sizes of routing tables are the
degrees of vertices, and stretches are quite small when the networks
show small-world properties. Comparing with previous work on
navigation \cite{Kleinberg2000,Watts2002,Krioukov2008}, our work may
provide profound insights into scalable routing scheme through a
self-organized way in the absence of prior knowledge.

\begin{acknowledgements}
This work is supported by the National Natural Science Foundation of
China under Grant Nos. 60874090, 60974079, 61004102. S-MC
appreciates the financial support of K.C. Wong Education Foundation,
China Postdoctoral Science Foundation, and Youth Innovation funding
of University of Science and Technology of China.
\end{acknowledgements}

\end{document}